\documentclass[preprint,12pt,authoryear]{elsarticle}

\usepackage{amssymb}
\usepackage{amsmath}

\usepackage{booktabs}
\usepackage{multirow}
\usepackage{subfigure}
\usepackage{afterpage}
\usepackage{natbib}
\usepackage[colorlinks=true, linkcolor=blue, citecolor=blue, urlcolor=blue]{hyperref}
\usepackage{adjustbox}
\usepackage{tabularx}
\usepackage{rotating}
\usepackage{graphicx}
\usepackage{float}
\usepackage[font=small]{caption}

\journal{}

\begin{document}

\begin{frontmatter}



\title{A Digital Twin Framework for Decision-Support and Optimization of EV Charging Infrastructure in Localized Urban Systems} 

\author[label1]{Bui Khanh Linh Do}
\author[label2]{Thanh H. Nguyen}
\author[label1]{Nghi Huynh Quang}
\author[label1,label3]{Doanh Nguyen-Ngoc\corref{cor1}}
\author[label1,label3]{Laurent El Ghaoui}

\affiliation[label1]{organization={Center for Environmental Intelligence, VinUniversity},
            city={Hanoi},
            country={Vietnam}}
            
\affiliation[label2]{organization={Department of Civil and Environmental Engineering, University of Illinois Urbana-Champaign},
            city={Urbana},
            state={IL},
            country={United States}}

\affiliation[label3]{organization={College of Engineering and Computer Science, VinUniversity},
            city={Hanoi},
            country={Vietnam}}
            
\cortext[cor1]{Corresponding author. E-mail address: doanh.nn@vinuni.edu.vn (Nguyen-Ngoc, D.).}

\begin{abstract}
As Electric Vehicle (EV) adoption accelerates in urban environments, optimizing charging infrastructure is vital for balancing user satisfaction, energy efficiency, and financial viability. This study advances beyond static models by proposing a digital twin framework that integrates agent-based decision support with embedded optimization to dynamically simulate EV charging behaviors, infrastructure layouts, and policy responses across scenarios. Applied to a localized urban site (a university campus) in Hanoi, Vietnam, the model evaluates operational policies, EV station configurations, and renewable energy sources. The interactive dashboard enables seasonal analysis, revealing a 20\% drop in solar efficiency from October to March, with wind power contributing under 5\% of demand, highlighting the need for adaptive energy management. Simulations show that dynamic notifications of newly available charging slots improve user satisfaction, while gasoline bans and idle fees enhance slot turnover with minimal added complexity. Embedded metaheuristic optimization identifies near-optimal mixes of fast (30kW) and standard (11kW) solar-powered chargers, balancing energy performance, profitability, and demand with high computational efficiency. This digital twin provides a flexible, computation-driven platform for EV infrastructure planning, with a transferable, modular design that enables seamless scaling from localized to city-wide urban contexts.

\end{abstract}



\end{frontmatter}



\section{Introduction}
\label{sec1}

As cities worldwide shift toward sustainable transportation, electric vehicles (EVs) and renewable energy integration have become essential in reducing carbon emissions and promoting environmental sustainability \citep{Al-Thani2022, Cao2021}. Global EV sales surged by 35\% in 2023, reaching 17 million projected by 2024 \citep{IEA2024}. While growth is strongest in China, the U.S., and Europe, emerging economies like Vietnam also experience rapid adoption, rising from fewer than 100 units in 2021 to over 87,000 by 2024. This transition necessitates a well-planned charging infrastructure \citep{Zhang2024, Yan2024}, yet deployment faces challenges such as increasing demand, inefficient station layouts, and the need to balance user satisfaction with grid stability \citep{Yu2021, Hussain2021}. Hybrid renewable systems incorporating solar and wind can enhance grid performance \citep{gorbunova-2020, Saadati2021} but require trade-offs between resilience, seasonal variability, and costs. Additionally, policies such as smart charging incentives and parking regulations influence EV adoption \citep{vanderKam2019, Mouhy-Ud-Din2024}, yet tools for policy evaluation remain scarce. The lack of integrated decision-support tools further limits the ability to optimize station deployment and policy effectiveness.

Existing studies rely heavily on static mathematical and optimization models that oversimplify EV behaviors and interactions with infrastructure \citep{Asna2021, Xiao2023}. Micro-level approaches, particularly agent-based modeling (ABM) using simulation software, have emerged to capture EV drivers’ daily travel patterns and charging behaviors \citep{Rasca2024, lopez-2021, marmaras-2017, novosel-2015, wang-2018}. However, most focus primarily on station placement and simple capacity sizing \citep{Cui2021, yi-2023}, often neglecting operational policies and energy variability. Many simulation frameworks also lack interactive dashboards, limiting practical application for planners \citep{Zeng2020}. These gaps highlight the potential of digital twin (DT) technology, which combines dynamic simulation with decision-support tools to optimize infrastructure under varying demand conditions. \citet{marcucci-2020} emphasized the need for integrating behavioral models within urban logistics DTs, while \citet{belfadel-2023} proposed a conceptual framework addressing key challenges in city logistics. However, the application of DTs for EV charging infrastructure planning remains largely unexplored.

This study addresses the following questions: (i) how can a digital twin with interactive simulation and visualization support EV infrastructure and energy planning? (ii) how can dynamic simulations and optimization algorithms be combined to evaluate policies and station configurations? We extend our prior work \citep{CSoNet_2024} by introducing a more comprehensive digital twin leveraging agent-based modeling and optimization algorithms for the optimal development of EV charging stations. This tool dynamically simulates EV-charging-energy interactions, offering visualization, scenario-based evaluation, and infrastructure optimization. Acting as a virtual laboratory, it enables stakeholders to test strategies before implementation, bridging the gap between simulation and practical infrastructure planning. While public charging networks are expanding, university campuses are emerging as critical hubs for EV adoption \citep{Caruso2017, Hovet2018}. This study applies the model to a university campus in Hanoi, Vietnam, named VinUniversity, where rising EV incentives and demand present challenges related to operational and infrastructure constraints. Although this study demonstrates the framework in a campus-scale environment, the modular structure and logic of the proposed digital twin approach are designed to scale effectively for urban system planning at district or city levels. This scalability enables the methodology and insights developed to support broader urban infrastructure planning, accommodating the complexities and dynamics of larger metropolitan areas.

\section{Literature review}
\label{sec2}

\subsection{Evaluating policy impact on EV Charging Station Operation}
\label{subsec2_1}

Policies play a critical role in optimizing EVCS utilization and reducing congestion. System dynamics models offer long-term insights into EV adoption and renewable energy incentives but often overlook short-term operational complexities. Agent-based simulations have evaluated policy effectiveness by modelling consumer behavior and competition between charging and fuel stations \citep{Esmaeili2022, Rasca2024}. \citet{li2-2024} demonstrated how agent-based simulation can assess urban policies such as access restrictions and shared mobility integration. However, most studies rely on predefined policies, limiting flexibility to combine or iteratively test strategies. Few decision-support tools assess how policies like smart charging incentives affect EV behavior and satisfaction \citep{vanderKam2019}. This highlights the need for a more flexible and comprehensive framework that integrates policy evaluation with infrastructure optimization, enabling data-driven decision-making in dynamic energy environments.

\subsection{EV charging infrastructure capacity optimization}
\label{subsec2_2}

Charging port numbers directly impact station throughput and user satisfaction. Queuing theory and heuristic algorithms, including Genetic Algorithms (GA), Particle Swarm Optimization (PSO), and Ant Colony Optimization (ACO), have been applied to solve complex location and sizing optimization problems, ensuring cost-effective and efficient EVCS deployment \citep{Asna2021, Xiao2023, Alhasan2023, Liu2012, Sun2022}. Agent-based modelling (ABM) has emerged as a dynamic alternative to static optimization approaches, providing a more detailed representation of user behavior and system interactions. While \citet{li-2021} revealed a lack of integrated agent-based simulations bridging urban mobility and logistics, and proposed a modular framework for future cross-sectoral modeling, \citet{aliedani-2018} demonstrated the potential of dynamic inter-agent communication through decentralized parking coordination, underscoring its value for improving urban mobility efficiency. ABMs simulate EV demand, charging station use, and grid impacts under varied conditions \citep{Mudiyanselage2024, Pagani2019, Cui2021, yi-2023} but rarely integrate operational policy or renewable energy considerations, failing to provide holistic EVCS planning, especially under large-scale adoption and energy transition.

\subsection{Planning renewable energy integration in EV charging stations}
\label{subsec2_3}

Hybrid renewable energy systems, combining solar and wind energy with battery storage, offer a promising approach to reducing grid dependency and enhancing charging station sustainability \citep{gorbunova-2020}. Optimization models suggest that integrating renewable energy into EV charging stations can enhance grid stability and accommodate more EVs \citep{Saadati2021}. Several studies have emphasized renewable energy investment as cost-minimization strategies \citep{Davidov2018, Li2022} but most assume average energy availability, missing seasonal variability. To address uncertainties in renewable energy supply, various optimization techniques have been applied. Nonlinear stochastic programming models, bilevel programming optimize EV charging behavior to mitigate energy intermittency and minimize costs \citep{Mehrjerdi2019, Zeng2020}. Beyond mathematical optimization, agent-based models are frequently employed to analyze the impact of energy policies and evaluate energy transition scenarios under uncertainty. Simulation tools such as MATLAB Simulink, HOMER, and GAMS are commonly used to optimize the performance of renewable energy-powered EV charging stations \citep{RaiHenry2016, Zeng2020, Diaz2024} but lack interactive dashboards, limiting their practicality for agile decision-making by non-technical stakeholders.

\subsection{Research gap and contribution}
\label{subsec2_4}

While significant progress has been made in EVCS optimization, key gaps remain. Static models oversimplify system dynamics and miss user interactions. Simulation-based approaches, particularly ABM, focus mainly on station placement while overlooking sizing, operational policies, and seasonal renewable energy effects. Most models also lack interactive dashboards, limiting practicality for decision-makers. Renewable energy integration remains underexplored, with many studies neglecting seasonal variations affecting energy availability and system performance. Additionally, optimization algorithms in EVCS planning often target station placement and operate in isolated analytical platforms, failing to offer integrated optimization within simulated realistic urban. This study addresses these limitations by developing a simulation-based decision-support tool integrating EV behavior modelling, policy evaluation, infrastructure optimization, and renewable energy management. The framework uses ABM to simulate user interactions and incorporates an interactive dashboard for scenario analysis. Embedded optimization algorithms within the digital twin environment support dynamic decision-making and efficient evaluation of planning strategies under varying demand and policy conditions. Bridging static modeling and practical decision-support, this tool provides a scalable, adaptable solution for optimizing EV charging networks.

\section{Methodology}
\label{sec3}

To support EV charging infrastructure planning and policy evaluation, this study develops an agent-based digital twin integrated with optimization and data analytics. Section \ref{subsec3_1} introduces the agent-based modeling framework, describing the design of agents, their interaction logic, as well as the input data and evaluation metrics. Section \ref{subsec3_2} outlines the setup for dynamic simulation experiments, statistical testing and the integration of optimization algorithms into the system. The complete agent-based simulation framework, including model structure, configuration files, simulation, and document (including appendix) is openly available at \href{https://github.com/dolinh11/EV-Solar-Sim}{GitHub}. The repository provides instructions for reproducing the scenarios and extending the model to other single-station contexts.

\subsection{Description of the Agent-Based Simulation Model}
\label{subsec3_1}

The agent-based model is implemented using the \href{https://gama-platform.org/}{GAMA platform}, an agent-oriented simulation platform inspired by object-oriented programming, particularly Java \citep{Taillandier2019, lesquoy2024gama}. GAMA was selected for its seamless integration of GIS data as the simulation environment and its ability to create interactive management dashboards with dynamic agent movement and visualizations, including various charts.

\subsubsection{Model environment}\label{subsubsec1}

In this study, we constructed the model environment using GIS layers from \href{www.openstreetmap.org}{OpenStreetMap} (OSM) and manually sketched building footprints from satellite imagery to represent roads, buildings, and infrastructure. The simulation incorporates spatial data to capture real-world layouts and support spatial allocation of agents and activities. Within the agent-based paradigm, the environment, including roads, buildings, and charging stations, is conceptualized as an active agent layer interacting dynamically with vehicle and energy agents. The space-time dynamic system allows agent decisions and system states to evolve simultaneously. The spatial layer can be easily modified or replaced, making the model applicable to different study areas. While currently applied to a localized site, the structure provides a transferable template for scaling to district or city-wide implementations without redesigning the agent logic.

\begin{figure}[H]
    \centering
    \includegraphics[width=0.75\linewidth]{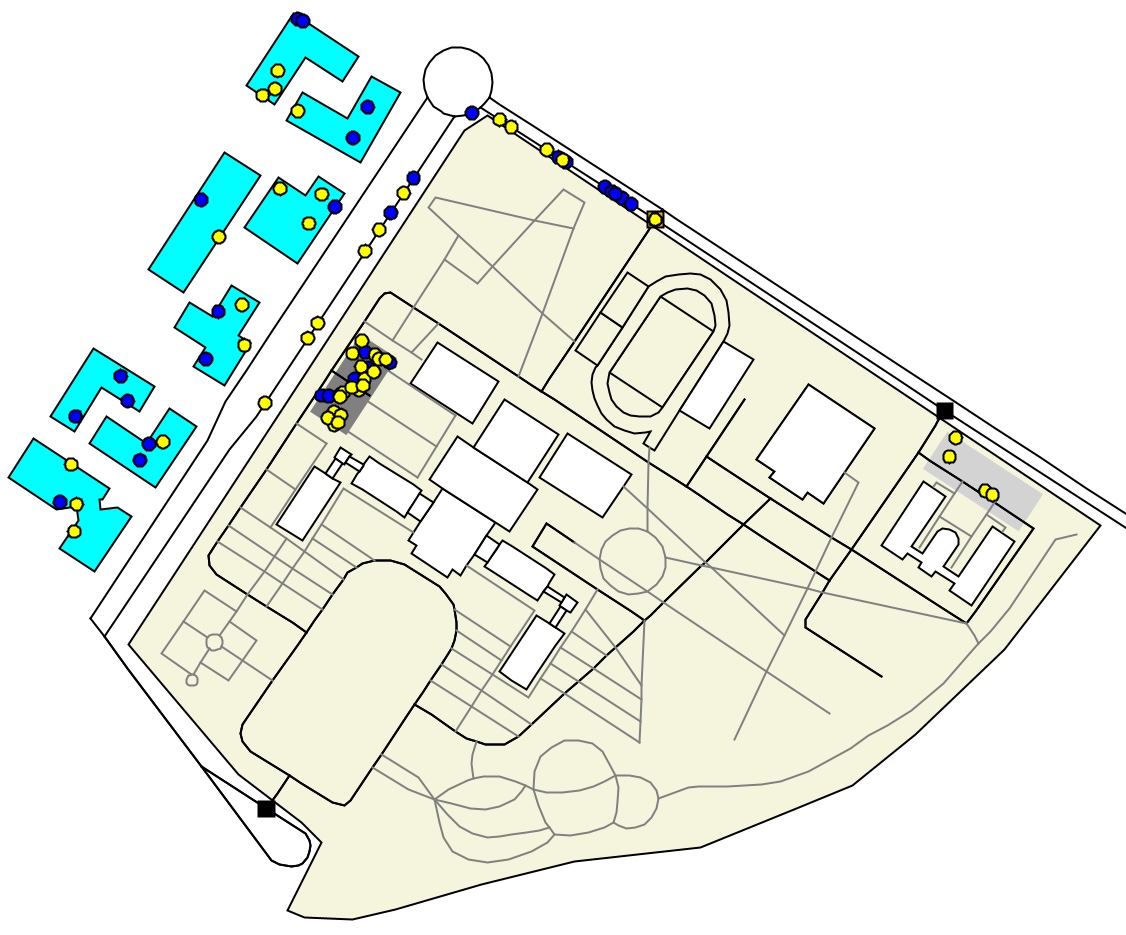}
    \caption{\footnotesize The model environment replicates the campus, manages vehicle flow, parking, and charging, and distinguishes vehicle agents: gasoline (blue circle) \& EVs (yellow circles).}
    \label{fig:map}
\end{figure}

Figure \ref{fig:map} illustrates the spatial data used, which primarily covers the VinUni campus (754m × 631m) and extends to nearby residential areas (1174m × 1131m) for user modeling. Key features include residential buildings (cyan) as vehicle home locations, parking areas (dark grey and light grey), and car-accessible roads (black lines) where interactions with moving agents occur. Other features, such as pedestrian-only paths (gray lines), campus boundaries (beige), and other campus buildings (white), are mainly for visualization. Three gates (diamond shapes) mark the main entry/exit points. OSM data enhances the simulation with geometric attributes like lane count, speed limits, and traffic direction, ensuring a realistic road network model \citep{Pham2020}. By integrating these elements, the model provides a comprehensive and realistic representation of the campus environment.

\subsubsection{Model entities}\label{subsubsec2}
The model consists of three interconnected sub-models: the charging station sub-model, the energy sub-model and the traffic sub-model. These components interact dynamically, reflecting the influence of EV charging behaviors, infrastructure configurations, and renewable energy conditions.

The charging station sub-model represents the campus environment, including buildings, roads, and charging areas. Residential buildings serve as the initialization points for vehicles, and the road network simulates real-world traffic infrastructure. Charging areas replicate the setup of parking area at building C (C-Parking) and building J (J-Parking) (dark grey and light grey rectangles in Figure \ref{fig:map}), with variables defining the number and type of charging ports (11kW and 30kW) as shown in Table \ref{table:CS_var}.

\begin{table}[h]
\caption{\footnotesize Variables defining the characteristics of charging areas agent, including the number and type of charging ports available at each location. }\label{table:CS_var}%
\scriptsize
\begin{tabular*}{\textwidth}{@{\extracolsep\fill}llll}
\toprule
\textbf{Variable Name}  & \textbf{Description}  & \textbf{Type}  & \textbf{Value Range} \\
\midrule
\texttt{location\_CS}          & The location of the charging area  & string   & \{"C-Parking", "J-Parking"\} \\
\texttt{activeCS\_11kW} & Number of 11kW CPs & integer  & [0; 50]  \\
\texttt{activeCS\_30kW} & Number of 30kW CPs  & integer  & [0; 10]  \\
\texttt{num\_CS}       & Total number of CPs at each building & integer  & [0; 100]  \\
\bottomrule
\end{tabular*}
\end{table}

The energy sub-model incorporates renewable energy sources into the charging infrastructure, consisting of three agents: BESS (Battery Energy Storage System), solarEnergy, and windEnergy. These agents compute energy production and system costs based on environmental factors such as solar irradiation, air temperature, and wind speed. Table \ref{table:energy_var} outlines the model’s key attributes, which serve as the primary parameters for energy output calculations, as used in \citep{Brihmat2014, Sun2020}.

\begin{table}[h]
\caption{\footnotesize Key attributes of agents in the energy sub-model, detailing parameters for energy output calculations based on renewable energy sources.}\label{table:energy_var}%
\scriptsize
\begin{tabular*}{\textwidth}{@{\extracolsep\fill}l l p{4cm} l p{2cm}}
\toprule
\textbf{Agent Name}  & \textbf{Variable Name}  & \textbf{Description}  & \textbf{Type}  & \textbf{Value Range} \\
\midrule
\multirow{2}{*}{\texttt{bess}}  
    & \texttt{bess\_capacity}  & BESS storage capacity  & float   & 80000 kW \\
    & \texttt{bess\_SoC}       & Stored energy in BESS  & float   & [0, 80000] \\
\midrule
\multirow{8}{*}{\texttt{solarEnergy}}  
    & \texttt{solar\_ghi}        & Solar irradiation  & float   & [0; 1000] \\
    & \texttt{air\_temp}         & Air temperature  & float   & [0; 40] \\
    & \texttt{nb\_solar}         & Number of PV panels  & integer  & [0; 1000] \\
    & \texttt{unit\_panel\_area} & Area of a PV panel  & float   & 2.7 m\textsuperscript{2} \\
    & \texttt{total\_panel\_area} & Total PV area  & float   & [0; 2700] \\
    & \texttt{current\_panel\_tmp} & Actual PV temperature  & float   & [0; 40] \\
    & \texttt{solar\_generated}   & Solar energy generated  & float   & [0; +$\infty$] \\
    & \texttt{solar\_cost}        & Total PV system cost  & float   & [0; +$\infty$] \\
\midrule
\multirow{4}{*}{\texttt{windEnergy}}  
    & \texttt{wind\_speed}      & Actual wind speed  & float   & [0; 10] \\
    & \texttt{nb\_wind}         & Number of turbines  & integer  & [0; 15] \\
    & \texttt{wind\_generated}  & Wind energy generated  & float   & [0; +$\infty$] \\
    & \texttt{wind\_cost}       & Total wind system cost  & float   & [0; +$\infty$] \\
\bottomrule
\end{tabular*}
\end{table}

The traffic sub-model governs vehicle movements, determining parking and charging slot selection. It includes two types of vehicle agents: gasoline-powered cars and electric vehicles, represented as blue and yellow circles, respectively, in Figure \ref{fig:map}. Table \ref{table:EV_var} shows EV agents' attributes such as state of charge (SoC), driver satisfaction, charging status, and station preferences \citep{vanderKam2019}. These agents make charging decisions based on station availability, energy demand, and individual priority, providing a realistic simulation of practical operational constraints.

To reflect the spatial dynamics of the study area, vehicle agents are initialized at designated residential buildings (represented by cyan shapes in Figure \ref{fig:map}), which serve as home locations outside the campus boundary. At the assigned start time, these agents are instantiated and move toward the three primary campus gates (represented by diamond shapes in Figure \ref{fig:map}) to enter the simulation environment. To maintain model generality and focus on aggregate charging demand, a unified empirical distribution is applied to all vehicle agents. This approach ensures the framework remains adaptable to various localized urban settings without requiring role-specific parameterization for different user groups, while still capturing the collective impact of campus traffic on charging infrastructure. The specific temporal logic governing when these agents transition from residential buildings to the campus environment is detailed in Section \ref{subsubsec3}.

\begin{table}[h]
\caption{\footnotesize Attributes for electric vehicles agent, including state of charge, owner satisfaction, charging status, and station preferences.}\label{table:EV_var}%
\scriptsize
\begin{tabular*}{\textwidth}{@{\extracolsep\fill}llll}
\toprule
\textbf{Variable Name}  & \textbf{Description}  & \textbf{Type}  & \textbf{Value Range} \\
\midrule
\texttt{nb\_electrical}  & Number of EVs per day. & integer  & [30; 200] \\
\texttt{is\_charging}    & EV is charging or not.  & bool  & \{true, false\} \\
\texttt{satisfied}       & EV can access a port when needed. & bool  & \{true, false\} \\
\texttt{SoC}            & Battery level (\%) assigned daily.  & float  & [20; 90] \\
\texttt{EV\_model}      & EV models at VinUni.  & string  & \{"VFe34", \newline "VF8", "VF9"\} \\
\texttt{priority\_des}  & Preference to stay in the current lot.  & bool  & \{true, false\} \\
\texttt{priority\_fast}  & Preference for fast/slow charging.  & bool  & \{true, false\} \\
\bottomrule
\end{tabular*}
\end{table}

This structure allows for an integrated, dynamic representation of campus transportation, energy usage, and charging behaviors.

\subsubsection{Model processes}\label{subsubsec3}

The simulation follows a time-stepped approach, with each step representing a five-minute interval to track vehicle movements and charging behaviors. If the charging infrastructure includes renewable energy, the energy sub-model is executed first to estimate daily energy generation. Then, the traffic and charging station sub-models are updated sequentially.

Weather data, including solar irradiation, air temperature, and wind speed, is retrieved at five-minute intervals at the beginning of each day, specifically for VinUni’s parking locations. Solar power generation follows the model from \citep{Brihmat2014}, where the output depends on solar irradiation and temperature, as shown in equation \eqref{eq:solar_power}. The photovoltaic (PV) power output \(P_{PV}(t)\) is calculated by:

\begin{equation}
P_{PV}(t) = k \cdot \frac{G(t)}{G_{ref}} P_{PV,STC} \cdot \eta_{PV} \cdot \left[1 - \beta_T \cdot (T_c(t) - T_{c,STC})\right] \label{eq:solar_power}
\end{equation}

where:
\begin{itemize}
    \item \(k = 1.15\): Bifacial PV panel absorption efficiency
    \item \(G(t)\): Solar irradiance at time \(t\), \(G_{ref} = 800\)W: Reference irradiance
    \item \(P_{PV,STC} = 610\)W: Rated power of PV panel
    \item \(\eta_{PV} = 22.6\%\): Solar absorption efficiency
    \item \(\beta_T = 0.0028\): Temperature-dependent power degradation coefficient.
    \item \(T_{c,STC} = 25\): Panel temperature under standard conditions.
    \item \(T_c(t)\): Panel temperatures computed by air temperature at time \(t\).
\end{itemize}

Wind power generation \(P_W(t)\) follows \citep{Sun2020}, calculated based on wind speed at the hub height, as shown in equation \eqref{eq:wind_power}.

\begin{equation}
P_W(t) = 
\begin{cases} 
0, & \text{if } v \leq v_{cut-in} \text{ or } v \geq v_{cut-out} \\
P_{rated} \cdot \frac{v^3(t) - v^3_{cut-in}}{v^3_r - v^3_{cut-in}}, & \text{if } v_{cut-in} \leq v \leq v_r \\
P_{rated}, & \text{if } v_r \leq v \leq v_{cut-out}
\end{cases} \label{eq:wind_power}
\end{equation}

where:
\begin{itemize}
    \item \(v(t)\): Actual wind speed at time \(t\), calculated proportionally based on actual wind speed and reference wind speed \& turbine height.
    \item \(P_{rated} = 3000\)W: Rated power output
    \item \(v_{cut-in} = 3.5\) m/s, \(v_{cut-out} = 45\) m/s: Cut-in and cut-out wind speeds
    \item \(v_r = 12\) m/s: Rated wind speed
\end{itemize}

\begin{figure}[H]
    \centering
    \includegraphics[width=0.9\linewidth]{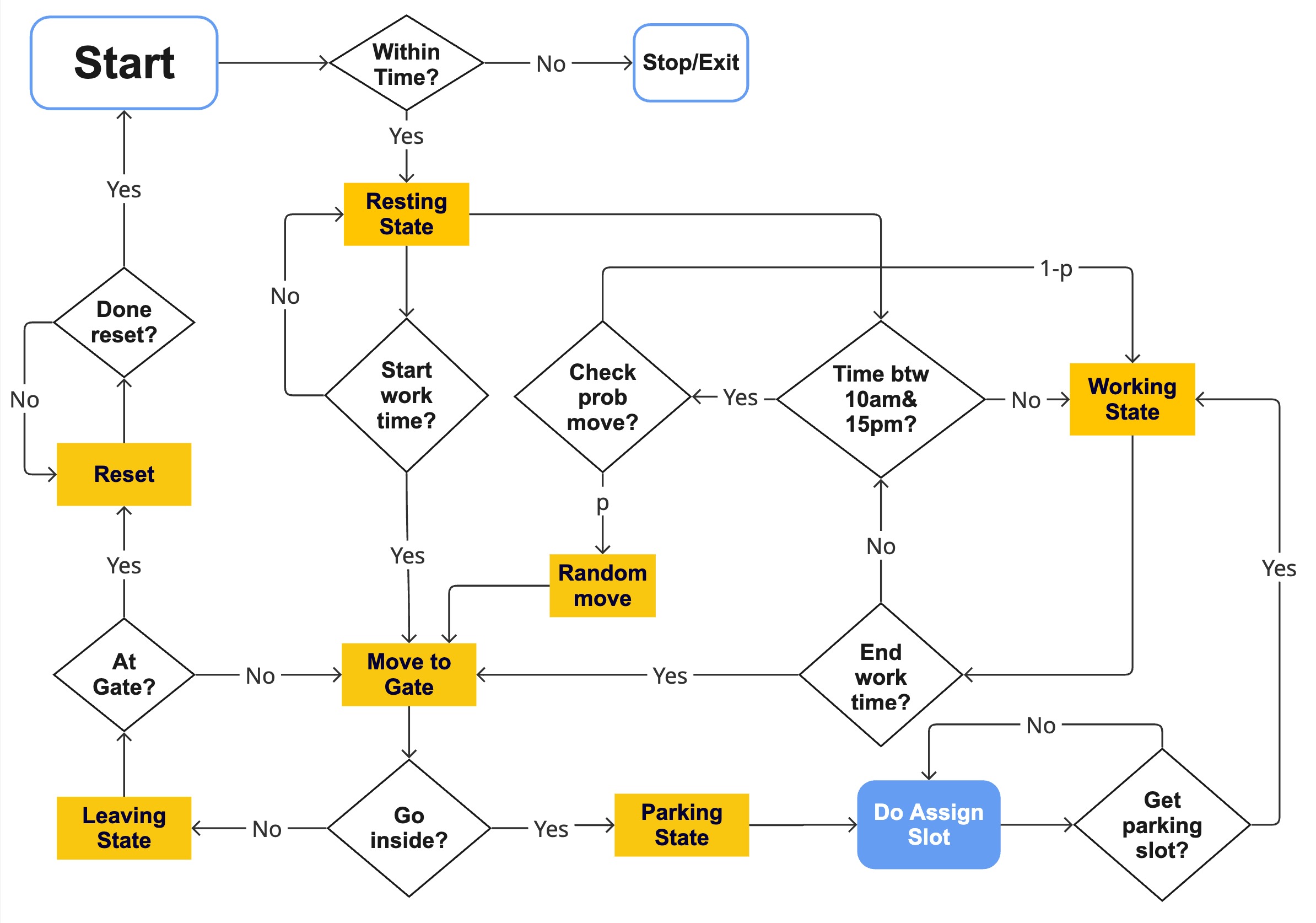}
    \caption{\footnotesize Stochastic decision-making framework for vehicle agents, featuring individualized schedules and probabilistic state transitions. Expanded from \citep{vanderKam2019, lee-2020}, the flowchart illustrates the daily cycle where arrival/departure times are stochastically sampled, and slot selection in "Do Assign Slot" is governed by SoC-based triggers and user preferences.}
    \label{fig:process_EV}
\end{figure}

The simulation is based on a dynamic simulation, a time-stepped approach where agents evolve continuously over time (5-minute resolution) according to predefined rules and events. At the start of each cycle, global variables are initialized (Table \ref{table:common_var}), and agents (e.g., vehicles) are instantiated to represent initial conditions. Each vehicle’s behavior is determined by its state (e.g., resting, parking, working, or leaving), with transitions triggered by temporal and probabilistic conditions. Figure \ref{fig:process_EV}, adapted and expanded from \citep{vanderKam2019, lee-2020}, illustrates the state transition process of the vehicle agent, detailing its decision-making in selecting parking and charging slots. Similar to prior studies that limited simulations to home–work travel due to simplifications and data constraints \citep{novosel-2015, lopez-2021}, our model also focuses on these two primary activity types to represent typical urban mobility relevant to EV charging.

\begin{table}[h]
\caption{\footnotesize Common state variables for electric and gasoline vehicles, outlining their parking and movement behaviors. }\label{table:common_var}%
\scriptsize
\begin{tabular*}{\textwidth}{@{\extracolsep\fill}llll}
\toprule
\textbf{Variable Name}  & \textbf{Description}  & \textbf{Type}  & \textbf{Value Range} \\
\midrule
\texttt{parking\_area}  & Targeted parking lot & chargingAreas  & \{"C-Parking", "J-Parking"\} \\
\texttt{home}  & Initial location & residential  & one\_of(residential\_areas) \\
\texttt{start\_work}  & Start work time & integer  & [8; 9] or [12; 14] \\
\texttt{end\_work}  & End work time & integer  & [17; 19] \\
\texttt{moving\_obj}  & Car’s movement state & string  & \{"resting", "parking", "working", "leaving"\} \\
\texttt{parking\_slot}  & Parking slot type & string  & \{"active\_CS", "inactive\_CS"\} \\
\texttt{in\_parkingArea}  & Car in parking area? & bool  & \{true, false\} \\
\bottomrule
\end{tabular*}
\end{table}

Vehicle behavior follows a daily cycle that begins in a Resting State. The arrival process at the campus gates is modeled as a time-dependent stochastic distribution informed by field observations. Rather than following a fixed schedule, the 'Start work time' for each agent is determined through a bimodal empirical sampling logic. Specifically, the framework utilizes configurable probability thresholds to allocate agents across observed peak windows: 70\% of agents are stochastically assigned a start time within the morning peak (8:00 - 9:00 AM), while the remaining 30\% are assigned to the early afternoon window (12:00 - 2:00 PM). Within these windows, individual arrival times are sampled using a Monte Carlo-based initialization, ensuring each agent possesses a unique schedule. Likewise, the 'End work time' is stochastically assigned to each agent within the observed window (5:00 – 7:00 PM), enabling the model to capture variability in user work durations and occupancy patterns. As shown in Figure \ref{fig:process_EV}, the 'Within Time?' node defines the boundary of this daily cycle. To capture observed non-stationary parking patterns, a mid-day turnover trigger is integrated between 10:00 AM and 3:00 PM, where agents have a 10–20\% probability of performing a random movement out of the campus.

A key distinction between gasoline vehicles and EVs lies in the "Do Assign Slot" function, which determines parking or charging slots based on the state of charge (SoC) and individual preferences. To preserve model generality and represent stochasticity in user behavior, the State of Charge (SoC) for each EV is randomized at the beginning of every simulation day, independent of the prior day’s travel history. This stochastic re-initialization of both agent schedules and initial SoC ensures behavioral heterogeneity and captures the dynamic variability inherent in charging demand. Such a simplification is methodologically necessary, as the precise estimation of daily trip distances and incremental energy depletion would necessitate continuous, high-resolution trajectory tracking, data that is currently unavailable in this specific urban context. Consequently, this design maintains robust dynamic variability across simulation cycles by focusing on the primary determinant of charging requirements: the SoC upon arrival. Vehicles with SoC $\leq$ 35\% prioritize charging while those with SoC between 35\% and 70\% decide to charge based on a predefined probability. Vehicles with SoC $\geq$ 70\% are considered sufficiently charged and are randomly assigned to inactive charging ports instead. Charging and station preferences (\textit{priority\_des} and \textit{priority\_fast}) influence slot selection.

The model continuously updates parameters like energy production, the SoC of EVs and traffic flow. At the end of the day, key performance indicators are aggregated and visualized to evaluate system performance and identify opportunities for optimization.

\subsubsection{Input Data}\label{subsubsec4}

A field study was conducted to analyze charging infrastructure and vehicle usage patterns in campus parking areas, including manual tracking of EVs and gasoline vehicles, EV type classification, and parking locations during peak hours (9:00 AM and 2:00 PM) over one week. These observations allowed for the generalization of vehicle movement patterns and the calibration of the model's stochastic parameters. Specifically, the observed arrival peaks justified the weighting for bimodal agent initialization. Given the university’s consistent schedule, these observations allowed generalization of vehicle movement patterns. Complementing this, an online survey and qualitative interviews focused on daily schedules, parking behavior, and EV charging habits. Most behavioral questions used a five-point Likert scale to assess user preferences. The combined dataset informed key behavioral parameters in the model, including arrival and departure distributions, random mid-day movements, SoC-based charging triggers (threshold on 35\%, 70\%), charging and parking priorities (90\% “Priority Destination” and 40\% “Priority Fast”), and low post-charging relocation tendencies. Together, these parameters generated realistic and dynamic parking and charging behaviors that underpinned the policy scenario setup. The model’s modular design ensures that such parameters can be easily calibrated for other single-station contexts (e.g., hospitals, residential complexes, commercial areas) across Vietnam, by updating the empirical distribution weights.

Energy consumption data for EV charging at C-Parking and J-Parking at VinUni’s total electricity usage (kWh) in 2023 were provided by the operational team. Historical solar and wind data from \href{https://solcast.com/}{Solcast}, validated against 48 BSRN ground stations, comprising over 1 million valid hourly average measurements and nearly 3 million instantaneous measurements \citep{Bright2019} were used for weather inputs in the simulation, ensuring accurate renewable energy integration.

\subsubsection{Evaluation metrics and objective functions}\label{subsubsec5}

The effectiveness of an EV charging station setup is assessed using three main metrics:  user satisfaction, renewable energy efficiency and financial performance.

User satisfaction: This metric represents the percentage of EVs that successfully complete charging, referred to as the satisfaction score. It is analogous to Station Utilization Rates, which assess charging station efficiency based on the number of charging requests served, as discussed in \citep{yi-2023, li-2024}. A higher satisfaction score indicates greater system efficiency, while lower values may signal issues such as energy shortages, congestion, or insufficient charging infrastructure. The satisfaction score is calculated as:

\begin{equation}
Satisfaction\ Score = \frac{EVs\ that\ completed\ charging}{EVs\ that\ requested\ charging}
\end{equation}

Renewable energy efficiency: This evaluates the relationship between energy generation, storage, and consumption, aiming to maximize self-consumption and reduce dependency on external energy \citep{merei-2016, Francisco2023, aldossary-2024}. Key ratios include:

\begin{equation}
    Self-Sufficiency = \frac{Consumed\ Renewable\ Energy}{Total\ Energy\ Demand}
\end{equation}

\begin{equation}
    Self-Consumption = \frac{Consumed\ Renewable\ Energy}{Total\ Generated\ Renewable\ Energy}
\end{equation}

Financial performance: This is analyzed through cost savings from renewable energy use and the payback period for investments \citep{merei-2016, Francisco2023}. A shorter payback period indicates greater economic feasibility. The initial renewable energy investment payback is calculated as:

\begin{equation}
Payback\ Period = \frac{Investment\ in\ Renewable\ System}{Monthly\ Profit}
\end{equation}

To ensure comparability, all values are normalized between 0 and 1. The payback period is normalized using the following function to prioritise profit return time within a predefined threshold of the payback period. This approach gives higher priority to payback periods within the acceptable range while reducing the weight of excessively long recovery times.

\begin{equation}
f(x) =
\begin{cases} 
\frac{x}{threshold}, & x \leq threshold \\
e^{\frac{threshold - x}{threshold}}, & x > threshold
\end{cases}
\end{equation}

Finally, the optimization metric combines the previous indicators to balance user satisfaction, energy self-sufficiency, and financial viability:

\begin{align}
    Objective\ Function = Satisfaction\ Score &+ Normalized\ Payback\ Period \notag \\
    &+ 0.8 \times Self\ Sufficiency
\end{align}

This linear function emphasizes maximizing user satisfaction and energy self-sufficiency while ensuring financial feasibility. The weight of 0.8 for self-sufficiency highlights the goal of integrating renewable energy as a supplementary, rather than exclusive, source.

\subsection{Experimental and Optimization Setup Description}
\label{subsec3_2}
\subsubsection{Dynamic Simulation Experiment Configuration}
For Policy Evaluation: The digital twin uses a batch experiment framework to evaluate various policy applications and their impact on user satisfaction. A four-tiered policy framework is tested to improve EV charging station operations:

\begin{itemize} 
    \item Policy 1: Ban gasoline vehicles from charging stations. 
    \item Policy 2: Impose 1,000VND/min idle fee after 30 minutes of full charge. 
    \item Policy 3: Relocate fully charged EVs to inactive stations. 
    \item Policy 4: Dynamic slot-availability notification system of charging ports.
\end{itemize}

In the current setup, user compliance with these four operational policies is modeled deterministically, assuming full adherence. This assumption is grounded in the real-world operational framework of Vietnam's charging network, where idle fees are automatically deducted through the official mobile app, and non-compliance can result in service restrictions (account suspension, wheel locking at selected sites, etc). Specifically, the dynamic notification system (Policy 4) represents an idealized information-sharing mechanism within the simulation environment.  It is intended to evaluate the potential impact of information transparency on user satisfaction rather than representing a live-streaming, sensor-based IoT deployment. While such a policy could be implemented in reality through mobile app integration, its role here is to support strategic scenario-based planning. To reflect potential behavioral variability, the model architecture allows future integration of a user compliance probability parameter for each EV agent. This allows stochastic reactions such as delayed or partial responses to system notifications or relocation requests.

\begin{table}[h]
    \centering
    \caption{\footnotesize Scenarios for implementing different operational policy combinations to evaluate their impact on EV charging station operations.}
    \label{tab:policy_cases}
    \begin{adjustbox}{width=\textwidth} 
    \begin{tabular}{lcccccc}
    \toprule
    \textbf{Policy / Case Level} & \textbf{Case 0} & \textbf{Case 1} & \textbf{Case 2} & \textbf{Case 3} & \textbf{Case 4} & \textbf{Case 5} \\
    \midrule
    Policy 1: Restrict gasoline vehicles & X & O & O & O & O & O \\
    Policy 2: Impose idle fee & X & X & O & O & O & O \\
    Policy 3: Relocate fully charged EVs & X & X & X & O & X & O \\
    Policy 4: Dynamic slot-availability notification & X & X & X & X & O & O \\
    \bottomrule
    \end{tabular}
    \end{adjustbox}
\end{table}

These policies are introduced incrementally across six scenarios (Table \ref{tab:policy_cases}), with varying EV numbers (50, 100, 150, 200), a fixed charging station setup including C-Parking (20×11kW, 4×30kW), J-Parking (15×11kW, 4×30kW), and 30 gasoline vehicles across all cases.

For Infrastructure Planning: To determine the optimal configuration for each EV demand scenario, the model simulates all possible parameter combinations across 4 scenarios of EVs number (50, 100, 150, 200). Table \ref{table:configuration} outlines the charging station setups and solar panel options, with a fixed BESS and an annual average weather dataset. Due to the expanding number of parameters, C-Parking with solar panels alone has 280 possible configurations, and including J-Parking increases this to 6,720. For simplicity, initial simulations focus on C-Parking, with optimization algorithms extending the analysis to J-Parking.

\begin{table}[h]
\centering
\caption{\footnotesize Value ranges for various configurations of charging stations and solar panels, including options for C-Parking and J-Parking.}
\label{table:configuration}
\scriptsize
\begin{tabular}{ll}
\toprule
\textbf{Category} & \textbf{Configuration Options} \\
\midrule
C-Parking & 11kW Charging Ports: 20, 25, 30, 35, 40, 45, 50 \\
          & 30kW Charging Ports: 2, 4, 6, 8, 10 \\
J-Parking & 11kW Charging Ports: 15, 18, 21, 24, 27, 30 \\
          & 30kW Charging Ports: 2, 4, 6, 8 \\
Solar Panels & 200, 300, 400, 500, 600, 700, 800, 900 \\
BESS capacity & 80kW\\
Weather Data & Annual average dataset \\
\bottomrule
\end{tabular}
\end{table}

\subsubsection{Statistical Testing for Dynamic Simulation with Wilcoxon Test}

The digital twin enables scenario-based simulations, producing time-series results that require statistical validation. Since these simulations often generate limited sample sizes and may involve non-normally distributed data, the Wilcoxon signed-rank test \citep{wilcoxon-1945} is an appropriate choice for comparing two paired datasets. Unlike parametric tests such as the paired t-test, which assumes normality, the Wilcoxon test is non-parametric and remains effective in handling skewed data and outliers.  

The test evaluates the null hypothesis (\(H_0\)) that the median difference between paired observations is zero. It ranks the absolute differences between paired samples, assigns ranks, and then calculates the test statistic \(W\) based on the sum of ranks.

\begin{equation}
W = \min(R^+, R^-)
\end{equation}

where \(R^+\) and \(R^-\) are the sums of positive and negative ranks, respectively. The result is then compared to critical values from the Wilcoxon distribution or assessed using a p-value to determine statistical significance. If \( p < 0.05 \), there is a statistically significant difference between the two scenarios, providing evidence to reject the null hypothesis \( H_0 \). Otherwise, if \( p \geq 0.05 \), no clear difference is detected between the two scenarios, meaning there is insufficient evidence to reject \( H_0 \). In this study, we employ the \texttt{scipy.stats.wilcoxon} function in Python, which automatically computes the test statistic and corresponding p-value for hypothesis testing.

\subsubsection{Optimization Algorithm Integration Settings}

The digital twin supports efficient decision-making by integrating various optimization algorithms (described in Table \ref{table:algorithms}), each employing a distinct approach to explore the solution space. Detailed pseudocode and parameter configurations (main parameters and initial values) for all six optimizers can be found in the Appendix on Optimization Algorithm in the accompanying GitHub repository.

\begin{table}[h]
\centering
\caption{\footnotesize Summary of optimization algorithms embedded in the digital twin, describing their approaches and objectives in exploring the solution space.}
\label{table:algorithms}
\scriptsize
\begin{tabular}{p{2.75cm} p{2.75cm} p{7cm}} 
\toprule
\textbf{Algorithm} & \textbf{Source} & \textbf{Description} \\
\midrule
Hill Climbing & \citep{pearl-1984} & A local search that iteratively improves the solution. \\
Simulated Annealing & \citep{kirkpatrick-1983} & A probabilistic technique that avoids local optima. \\
Tabu Search & \citep{glover-1989, glover-1990} & Uses memory structures to avoid revisiting solutions. \\
Reactive Tabu Search & \citep{battiti-1994} & Adjusts memory size for improved efficiency. \\
Genetic Algorithm & \citep{holland-1975, goldberg-1988} & A population-based algorithm inspired by evolution. \\
Particle Swarm Optimization & \citep{Kennedy1995} & A swarm intelligence method where particles adjust positions based on experience. \\
\bottomrule
\end{tabular}
\end{table}

These algorithms are assessed based on their accuracy relative to full-case results and runtime, using the Normalized Euclidean Distance (NED) as a comparative metric to quantify deviations from a reference solution, ensuring fair comparison across multiple parameters. Given two solutions \(\mathbf{x}_1 = (x_{1,1}, x_{1,2}, \dots, x_{1,n})^\top\) and \(\mathbf{x}_2 = (x_{2,1}, x_{2,2}, \dots, x_{2,n})^\top\), NED is computed as equation \ref{eq:NED}, where \(s_j\) represents the feasible range of the \(j\)-th variable, ensuring consistency across different scales.

\begin{equation}
D = \sqrt{\sum_{j=1}^{n} \left( \frac{x_{1,j} - x_{2,j}}{s_j} \right)^2 }
\label{eq:NED}    
\end{equation}

By applying NED, we obtain a standardized accuracy measure, enabling fair performance assessment across optimization approaches. In this study, NED thresholds are analyzed based on realistic parameter ranges, such as solar panel deployment and charging infrastructure capacity.

\section{Results}
\label{sec4}

\subsection{Digital Twin enables description and monitoring of EV charging infrastructure}
\label{subsec4_1}

This section describes the interactive dashboard featured in our digital twin, highlighting a use case where the dashboard is utilized to evaluate seasonal solar energy fluctuations.
\subsubsection{Dashboard provides dynamic map and monitoring charts}

\begin{figure}[H]
    \centering
    \includegraphics[width=\linewidth]{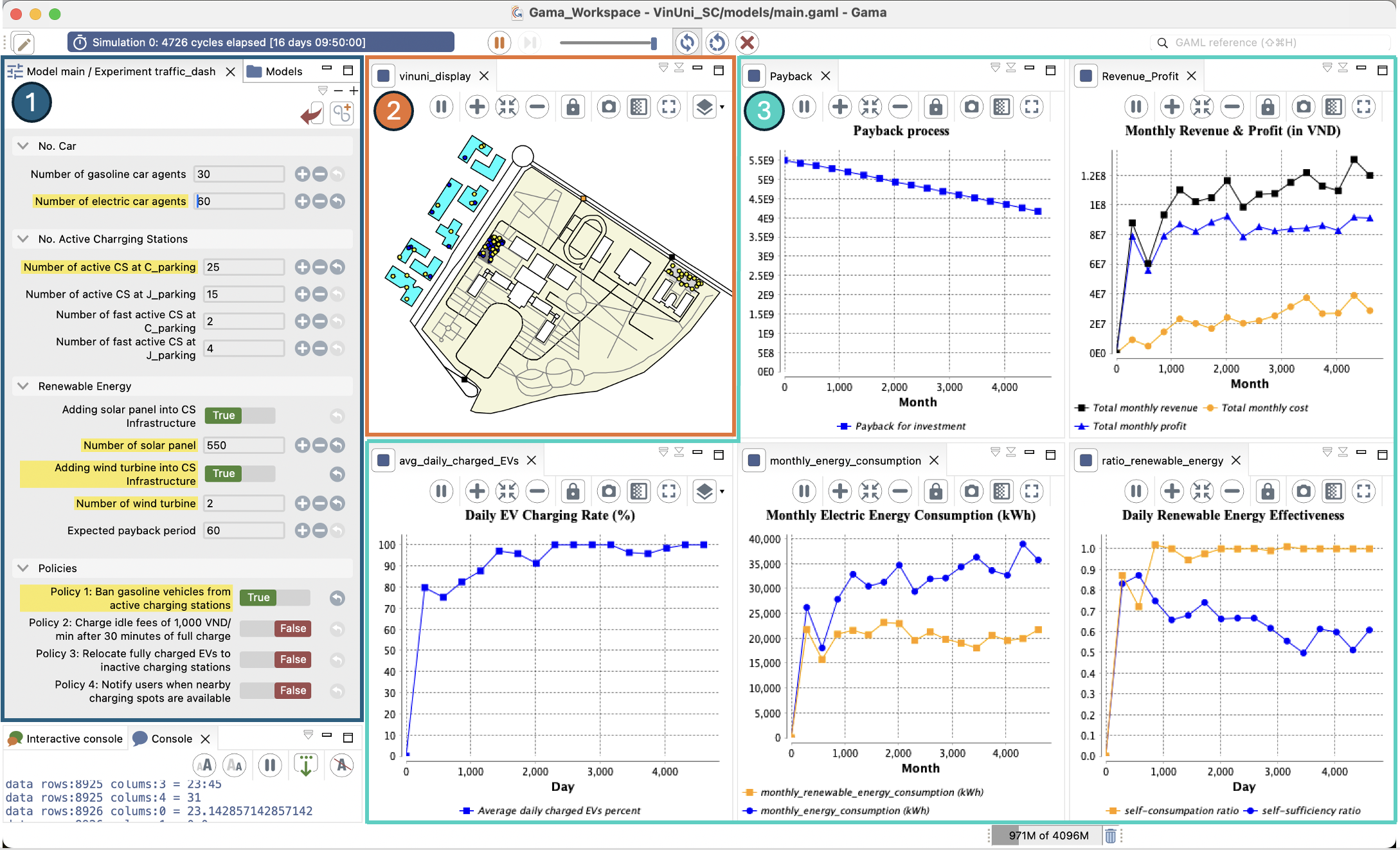}
    \caption{\footnotesize Digital Twin dashboard for  EV charging visualization, integrating (1) Interactive parameters control panel, (2) GIS-based interactive multi-agent simulation, and (3) Performance charts tracking user satisfaction, energy efficiency, and financial viability.}
    \label{fig:dashboard}
\end{figure}

Our dashboard, shown in Figure \ref{fig:dashboard}, provides a dynamic visualization of EV charging operations by integrating an interactive map and monitoring charts that allow users to dynamically explore charging scenarios, adjust station configurations, and evaluate policy impacts. The interface consists of three main components: (1) an interactive control panel (left-hand side) for adjusting station parameters and operational policies, (2) a map-based view (center) displaying the current status of charging stations, and (3) a data analytics section presenting key performance indicators (KPIs) such as user satisfaction, energy efficiency, and financial viability.

The focus on interaction within the simulation environment is intentional, as it enables planners to intuitively explore and compare infrastructure or policy scenarios under evolving local conditions. The dashboard provides an interactive interface where users can visualize how changes in charging policies, station capacities, or demand affect system indicators such as satisfaction, SoC evolution, and energy balance. The system thereby supports scenario-based decision-making, allowing users to simulate new policy or infrastructure configurations, observe outcomes through live-updating maps and charts, and assess their viability before implementation. This interactive setup facilitates rapid scenario evaluation and enhances understanding of system behavior, which is especially valuable in emerging EV markets like Vietnam, where data streams for continuous live operational control are still limited but scenario-based decision tools are urgently needed. Operators can proactively manage infrastructure, optimize resources, and uncover hidden demand patterns for future planning.

\subsubsection{Dashboard reveals seasonal solar fluctuations' impact on energy efficiency and financial viability}

To assess the impact of seasonal fluctuations on renewable energy efficiency, a one-month dynamic simulation was conducted with 50 EVs and 500 solar panels, keeping other station configurations constant. Four distinct weather datasets, each representing a quarterly average, were used to simulate different seasonal conditions.

\begin{figure}[H]
    \centering
    \includegraphics[width=\linewidth]{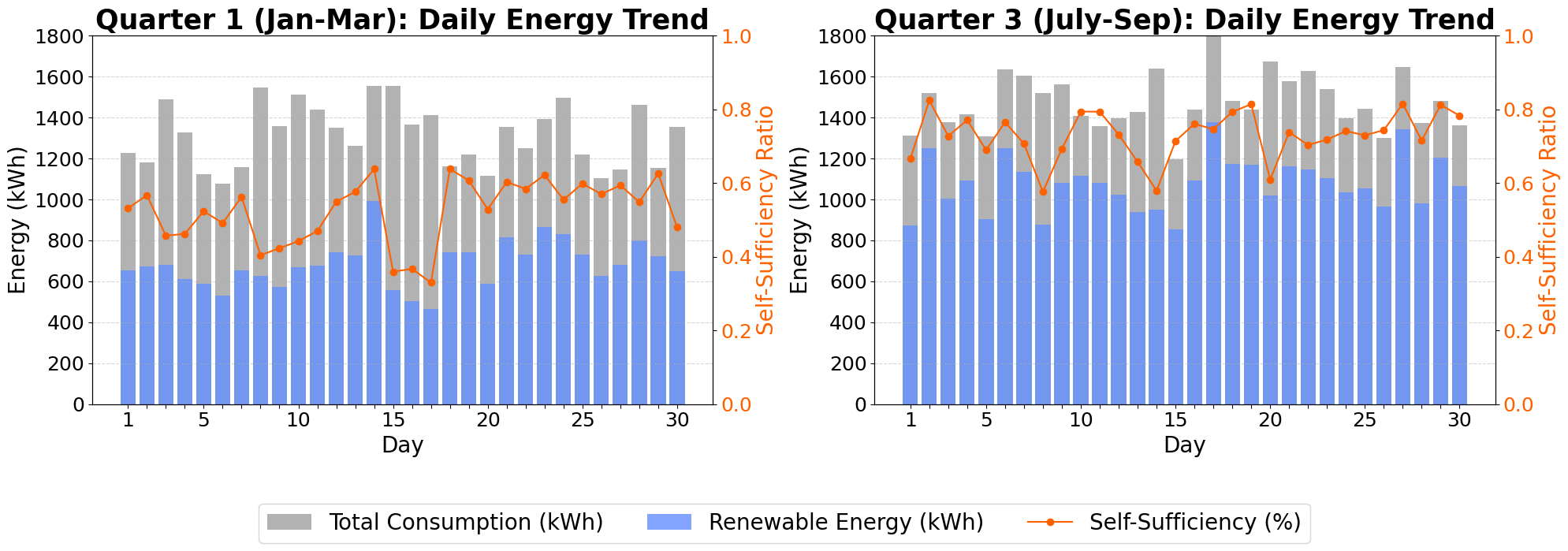}
    \caption{\footnotesize Average one-month simulation results for Quarter 1 \& Quarter 3, showing total energy consumption (kWh), renewable energy usage (kWh), and self-sufficiency (\%) without wind energy integration.}
    \label{fig:quarter_sim}
\end{figure}

As shown in Figure \ref{fig:quarter_sim}, results indicate that solar energy output peaks from July to September (Quarter 3), covering 73\% of daily electricity demand (1,078 kWh/day). In contrast, from January to March  (Quarter 1), solar output declined by 21\%, meeting only 52\% of demand (681 kWh/day). This seasonal fluctuation significantly affects self-sufficiency and financial returns. During July to September, higher solar generation reduces reliance on grid electricity, leading to a monthly profit of 100 million VND, 43\% higher than in January to March, when lower solar availability increases grid dependency, limiting profit to 70 million VND despite similar demand conditions.

\begin{figure}[H]
    \centering
    \includegraphics[width=\linewidth]{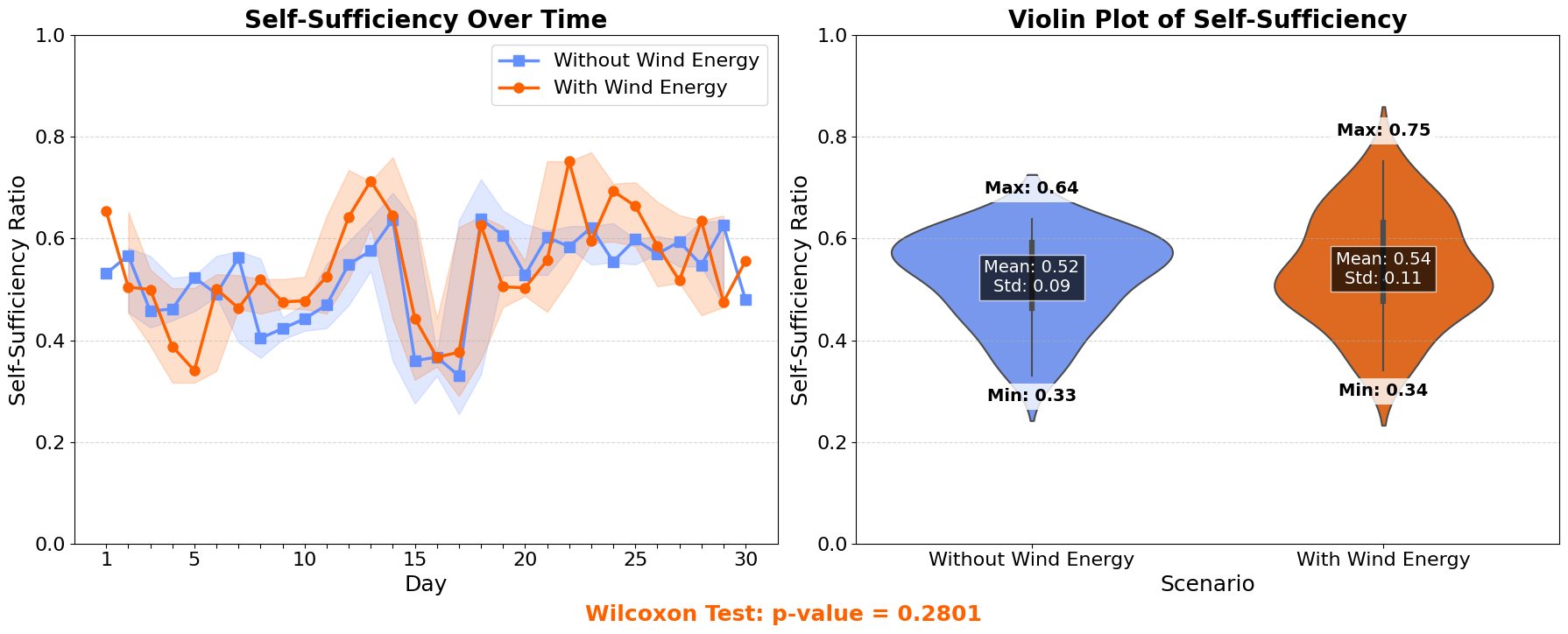}
    \caption{\footnotesize Average one-month simulation results for Q1, comparing self-sufficiency (\%) with (orange) and without (blue) wind energy integration using Wilcoxon Test.}
    \label{fig:quarter_sim_wind}
\end{figure}

Wind energy was assessed as a potential complement to solar power during October to March, when solar output was lower. A one-month simulation with 20 wind turbines at 100m height (Figure \ref{fig:quarter_sim_wind}) showed a marginal improvement in self-sufficiency, with the mean self-sufficiency ratio increasing slightly from 0.52 (SD = 0.09) to 0.54 (SD = 0.11). The Wilcoxon test (p = 0.2801) confirmed that this difference was statistically insignificant. Consequently, mean daily renewable energy consumption rose only modestly, from 681 kWh to 718 kWh. The limited effectiveness is attributed to peak wind generation occurring outside primary charging hours, specifically from 7:00 PM to 6:00 AM, reducing its direct usability. Despite requiring a 28\% increase in initial investment, wind energy contributes under 5\% to the total energy supply, but further extending the return-on-investment period.

To better utilize solar energy, prioritizing BESS expansion over wind investment would be more effective, given the weak wind energy potential in Vietnam. While the simulation includes an 80 kW BESS, further optimization strategies could enhance its efficiency, particularly in scenarios with dynamic electricity pricing and adaptive energy management.

\subsection{Dynamic simulations systematically evaluate operational policies and infrastructure expansion}
\label{subsec4_2}

Section 4.2 discusses dynamic simulations to explore the effects of operational policies, evaluate charging configurations, and conduct a sensitivity analysis to identify key parameters for optimizing EV charging infrastructure.

\subsubsection{Notification systems improve satisfaction, while idle fees increase slot availability with simple enforcement}

\begin{figure}[H]
    \centering
    \includegraphics[width=0.7\linewidth]{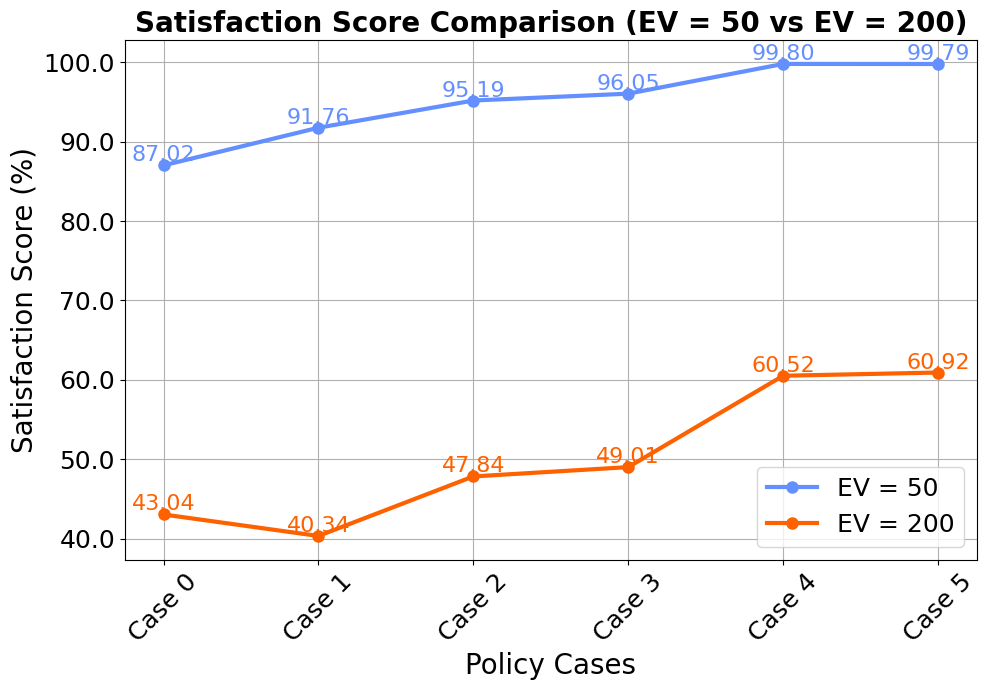}
    \caption{\footnotesize EV satisfaction scores for six combinations of four policy interventions, including gasoline bans, idle fees, EV relocation, and dynamic slot-availability notification (defined in Table \ref{tab:policy_cases}).
}
    \label{fig:policy_v1}
\end{figure}

Simulation results highlight the effectiveness of four key policy interventions, including banning gasoline vehicles, imposing idle fees, relocating fully charged EVs, and implementing a notification system, in improving EV user satisfaction (Figure \ref{fig:policy_v1}). In the 50-EV scenario, without intervention, gasoline vehicles occupy charging slots, and unregulated EV parking reduces station turnover, keeping satisfaction below 90\%. Introducing a gasoline vehicle ban (Case 1) increases station accessibility, while implementing an idle fee (Case 2) further optimizes utilization, raising satisfaction to 95\%. However, enforcing EV relocation (Case 3) provides minimal additional benefit (less than 1\% increase), indicating that this measure has limited impact. The notification system (Case 4, 5) proves to be the most effective intervention, dynamically informing users of available slots and boosting satisfaction by 5\% to nearly 100\% in low-demand scenarios (50 EVs). In high-demand cases (200 EVs),  the overall trend remains consistent, and the notification system has an even greater effect, increasing satisfaction by over 10\% compared to other policies, reaching 60\%. However, deploying such a system introduces added complexity and costs, which must be carefully weighed against its benefits.

\begin{figure}[H]
    \centering
    \includegraphics[width=0.7\linewidth]{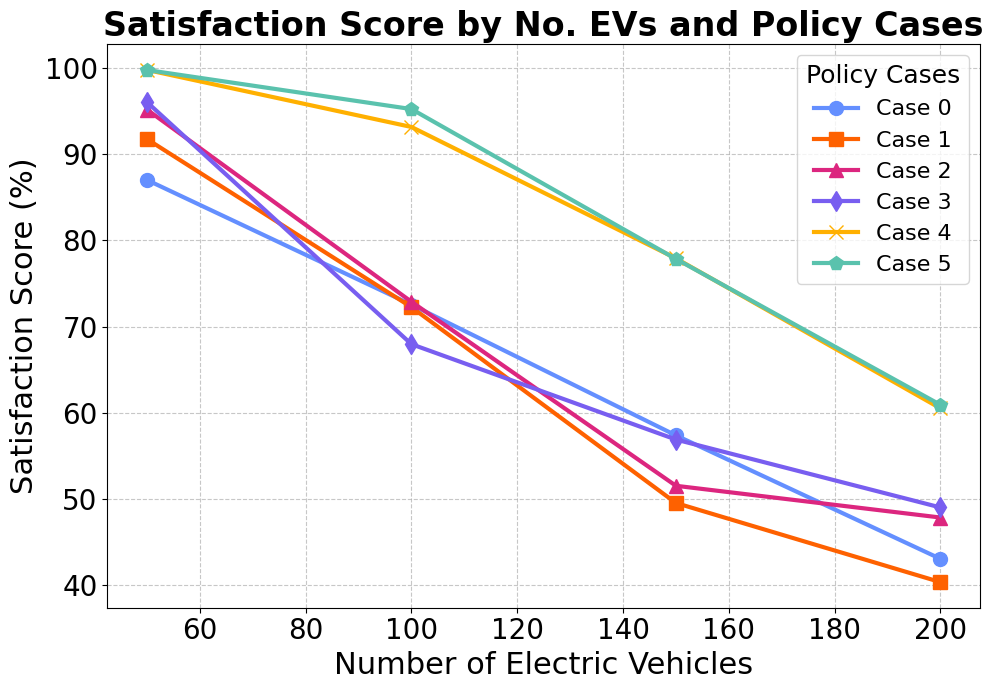}
    \caption{\footnotesize EV satisfaction scores across six policy combination cases (defined in Table \ref{tab:policy_cases}) as EV size increases from 50 to 200.}
    \label{fig:policy_v2}
\end{figure}

Despite these improvements, policies alone cannot fully compensate for infrastructure limitations (Figure \ref{fig:policy_v2}). Satisfaction drops sharply as EV numbers increase, declining from 87\% at 50 EVs to just 43\% at 200 EVs, even with multiple policies in place. These findings indicate that while policy interventions enhance station efficiency, they must be complemented by infrastructure expansion to sustain long-term EV adoption. The most effective approach involves banning gasoline vehicles and imposing idle fees to enhance slot turnover, implementing a dynamic slot-availability notification system for slot allocation despite its complexity, and scaling charging infrastructure to accommodate growing demand.

\subsubsection{Dynamic simulation identifies optimal configurations for 30kW and 11kW chargers and solar energy}

\label{subsubsec4_2}

\begin{figure}[H]
    \centering
    \includegraphics[width=\linewidth]{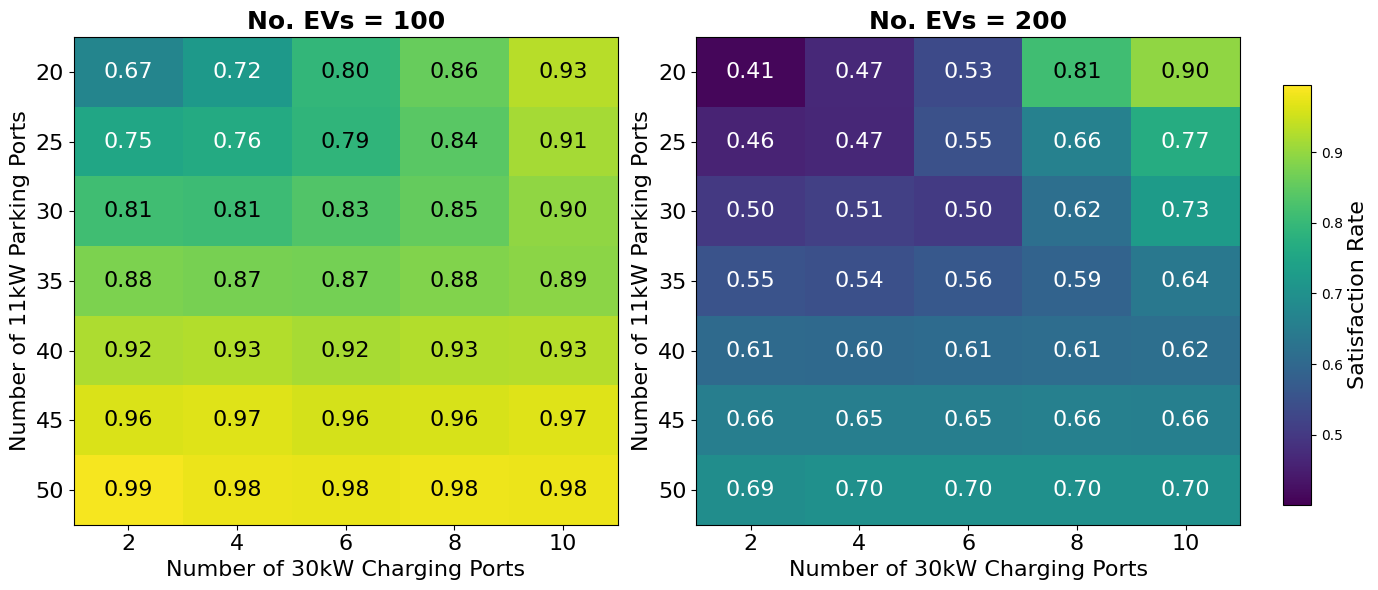}
    \caption{\footnotesize Satisfaction scores for different 30kW (horizontal axis) and 11kW (vertical axis) charger combinations under two demand scenarios: 100 EVs (left) and 200 EVs (right).}
    \label{fig:heatmap_CS}
\end{figure}

Results from the 280 configuration experiments (Figure \ref{fig:heatmap_CS}) reveal distinct performance trends across EV demand levels. For lower demand (50–100 EVs), increasing 11kW chargers steadily improves satisfaction, reaching nearly 100\% at 30 chargers (50 EVs) and 50 chargers (100 EVs). However, in higher demand scenarios (150–200 EVs), fast chargers (30kW) become more critical, significantly enhancing satisfaction compared to adding more 11kW units. Satisfaction peaked with 20 (11kW) and 10 (30kW) chargers. Adding more than 40 (11kW) units resulted in diminishing returns, as congestion became uneven, some chargers were underused while others were overloaded. These findings emphasize the need for a balanced mix of slow and fast chargers to maximize station efficiency.

\begin{figure}[H]
    \centering
    \includegraphics[width=0.7\linewidth]{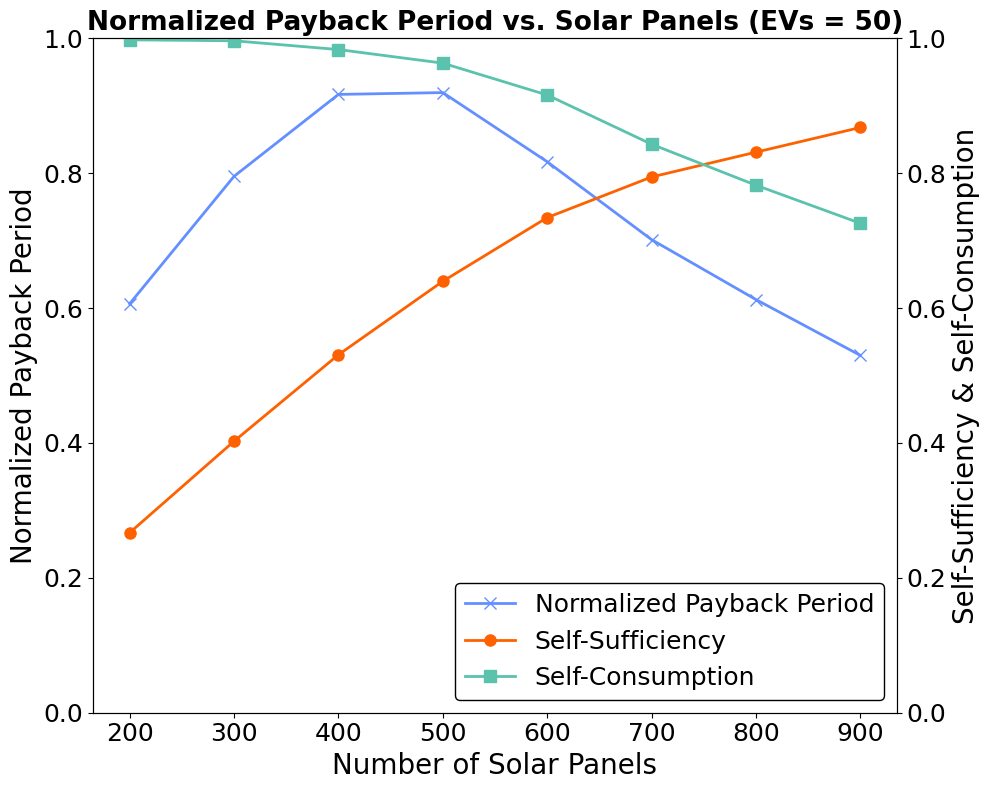}
    \caption{\footnotesize Relationship between number of solar panels and two key metrics: the normalized payback period (blue) and the self-sufficiency and self-consumption (orange and green, respectively) for the 50 EVs scenario.}
    \label{fig:line_CS}
\end{figure}

Regarding solar panel integration in the 50-EV scenario, Figure \ref{fig:line_CS} shows that self-consumption remains consistently high ($\ge 80\%$), indicating efficient solar energy utilization. However, self-sufficiency starts below 30\% but rises to nearly 90\% as panel capacity increases. Despite this improvement, economic trade-offs emerge: increasing solar capacity reduces payback efficiency, delaying profitability. With a payback period threshold of 60 months, the results demonstrate that beyond 500 solar panels, additional installations offer negligible financial returns, indicating that oversizing the PV system is not cost-effective. Similar trends are observed across all EV demand scenarios, reinforcing the importance of aligning solar deployment with charging station capacity to optimize investment.

The optimal configurations derived from the Objective Function evaluation (Table \ref{tab:full_case}) balance these trade-offs, ensuring high satisfaction, cost efficiency, and energy performance. The results underscore the importance of a well-balanced mix of 30kW and 11kW chargers, along with an optimally sized solar array, to maximize station efficiency and economic sustainability.

\begin{table}[h]
\centering
\caption{\footnotesize Results from the full case scenario with 280 combinations, detailing the optimal setups for four electric vehicle demand scenarios (50, 100, 150, 200 EVs). \newline
\textit{Note}: CS = Charging Station; CPs = Charging Ports.}
\label{tab:full_case}
\scriptsize
\begin{tabular}{lcccc}
\toprule
\textbf{} & \textbf{50 EVs} & \textbf{100 EVs} & \textbf{150 EVs} & \textbf{200 EVs} \\
\midrule
\textcolor{red}{\textbf{CS Configuration}} \\
No. CPs 11kW          & 35  & 50  & 20  & 20  \\
No. CPs 30kW          & 4   & 4   & 10  & 10  \\
No. Solar Panels      & 500 & 900 & 600 & 600 \\
\midrule
\textbf{Evaluation Metrics} \\
Satisfaction Score     & 1.00 & 0.99 & 0.94 & 0.89 \\
Self-Consumption      & 0.99 & 0.98 & 0.95 & 0.93 \\
Self-Sufficiency      & 0.67 & 0.60 & 0.59 & 0.57 \\
Normalized Payback    & 0.94 & 0.98 & 0.95 & 0.92 \\
Objective Function    & 2.47 & 2.45 & 2.37 & 2.26 \\
\bottomrule
\end{tabular}
\end{table}

\subsubsection{Sensitivity analysis highlights 30kW chargers and notifications boost satisfaction, while solar panels drive self-sufficiency and payback}

\begin{figure}[H]
    \centering
    \includegraphics[width=0.9\linewidth]{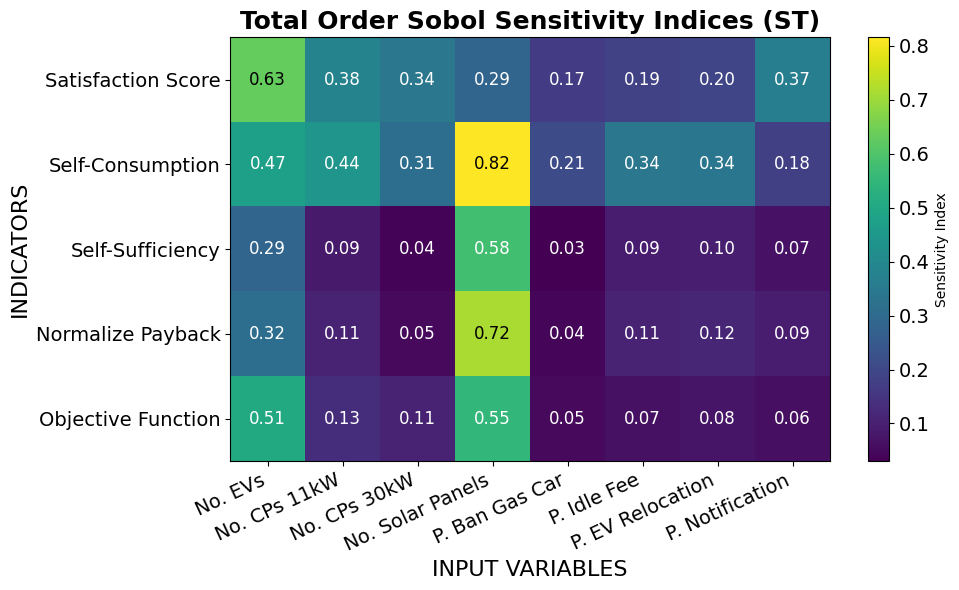}
    \caption{\footnotesize Heatmap of the Total-Order Sobol Sensitivity Indices (ST) of key factors affecting EV satisfaction, station efficiency, and financial viability.}
    \label{fig:sobol_v2}
\end{figure}

Sensitivity analysis identifies the most influential factors affecting EV satisfaction, station efficiency, and financial viability. As shown in Figure \ref{fig:sobol_v2}, results indicate that fast charger availability (30kW) and the implementation of a notification system are the two most critical variables impacting satisfaction scores. The Total-Order Sobol Indices reveal that EV satisfaction is primarily influenced by the number of 11kW (ST = 0.38) and 30kW chargers (ST = 0.34), with only a minor difference of 0.04. Despite their smaller range and step size (2–10 with step 2 vs. 20–50 with step 5 for 11kW chargers), 30kW chargers demonstrate a comparable impact, reinforcing their efficiency in high-demand scenarios. Among policy interventions, the notification system (ST = 0.37) has the most substantial effect, dynamically improving slot allocation and reducing waiting times.

For energy performance, solar panel capacity has the strongest influence on self-sufficiency and payback period. Additionally, idle fees and relocation policies (both ST = 0.34) significantly enhance self-consumption by optimizing station utilization and aligning charging demand with solar availability. However, since their effects are nearly identical, forced relocation may offer little added value over idle fees, questioning its necessity given the additional enforcement effort required.

These results align and reinforce the conclusions from Sections \ref{subsec4_2}, emphasizing the critical role of fast chargers, notification systems, and optimal solar panel sizing in improving EV charging efficiency and financial sustainability.

\subsection{Embedded optimization algorithms in digital twin enable efficient decision-making}
\label{subsec4_3}
Our Digital Twin tool provides a comprehensive optimization environment, enabling both exhaustive batch experiments for small-scale scenarios and embedded optimization algorithms for efficiently finding near-optimal solutions in complex, high-dimensional cases. By integrating optimization directly into the Digital Twin framework, our tool significantly reduces computational time while maintaining solution quality.

As shown in Section \ref{subsubsec4_2}, exhaustive batch simulations allow for evaluating all possible configurations, providing key insights into optimal infrastructure setups, variable impacts, and system patterns. However, as the number of parameters increases, the computational burden grows exponentially, making exhaustive testing impractical. To address this, our tool directly embeds optimization algorithms within the simulation environment, unlike other studies that require external execution. Six widely used algorithms, hill climbing, simulated annealing, tabu search, reactive tabu search, genetic algorithm, and particle swarm optimization (PSO), have been integrated to efficiently explore the solution space and identify near-optimal configurations for large-scale decision-making.

\begin{figure}
    \centering
    \includegraphics[width=0.8\linewidth]{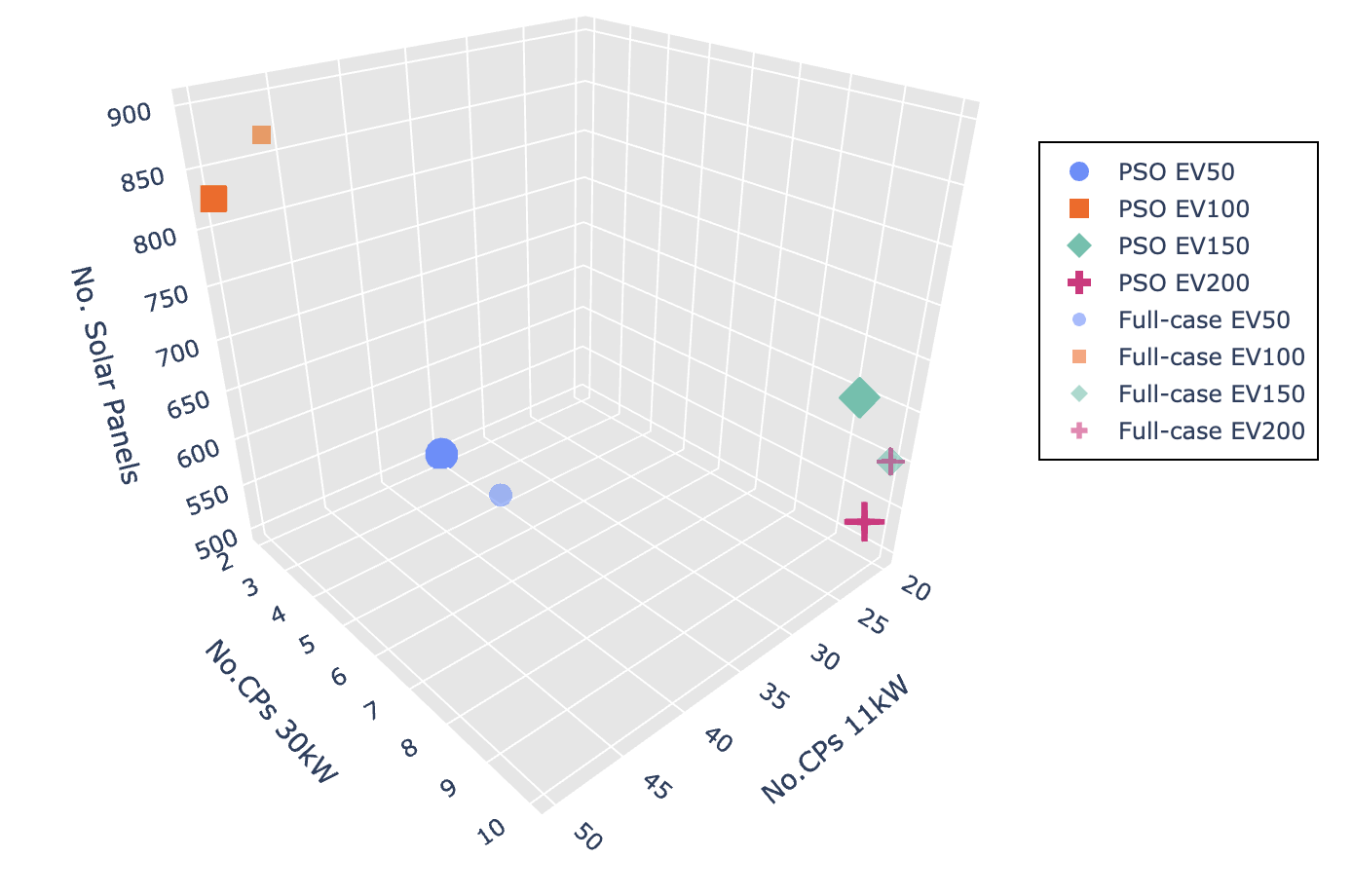}
    \caption{\footnotesize Comparison of PSO optimization vs. full-case results across solar panels and charging port configurations (11 kW \& 30 kW). Each point represents configurations for 50 EVs, 100 EVs, 150 EVs, and 200 EVs, with PSO results shown as bigger markers.}
    \label{fig:opt_res}
\end{figure}

The optimization algorithms generate solutions closely aligned with full-case results, with minor deviations (at most one adjustment step): at most $\pm$5 charging ports (11kW), $\pm$2 charging ports (30kW), and $\pm$100 solar panels. To illustrate the efficiency gains, we analyze a representative optimization run using PSO, as shown in Figure~\ref{fig:opt_res} and Table~\ref{tab:opt_res}. The Normalized Euclidean Distance values (ranging from 0.656 to 1.345) indicate minimal deviation between PSO and full-case solutions. For EV50 and EV200, distances of 1 and 0.656 show near-identical results. Slightly higher values for EV100 and EV150 (1.238 and 1.345) stem from the relative influence of solar panel quantities in the optimization process, as they contribute more significantly to overall system performance compared to charging port variations. Since the full-case method only considers predefined discrete values, while PSO can optimize outside these constraints, some deviation is expected. However, the results confirm that PSO delivers near-optimal solutions while reducing computational time by 82\%, as PSO only needs to evaluate around 50 configurations compared to the 280 combinations required in the full-case scenario.

\begin{table}[h]
\centering
\caption{\footnotesize Results comparison of PSO with full-case, highlighting configurations for different EV cases and their normalized Euclidean distances.}
\label{tab:opt_res}
\scriptsize
\setlength{\tabcolsep}{4pt} 
\renewcommand{\arraystretch}{1.2} 
\begin{tabular}{lcccc}
\toprule
\textbf{EV Case} & \multicolumn{2}{c}{\textbf{No. 11kW, No. 30kW, No. Solar}} & \textbf{Norm. Euclidean Distance} \\
\cmidrule(lr){2-3}
& \textbf{PSO} & \textbf{Full-case} &  \\
\midrule
EV50  & (35, 2, 500)  & (35, 4, 500)  & 1.000 \\
EV100 & (50, 2, 827)  & (50, 4, 900)  & 1.238 \\
EV150 & (25, 10, 690) & (20, 10, 600) & 1.345 \\
EV200 & (22, 10, 548) & (20, 10, 600) & 0.656 \\
\bottomrule
\end{tabular}
\end{table}

Finally, applying PSO to optimize charging configurations for both C-Parking and J-Parking, which involve 6,720 combinations, 24 times more than cases without J-Parking variations, yields results after evaluating only about 50 configurations, as shown in Table \ref{tab:Jpaking_res}. This optimization determines the optimal allocation of charging ports (11kW \& 30kW) and solar panels for both locations.

\begin{table}[h]
\centering
\caption{\footnotesize Optimal setups determined by PSO for both C-Parking and J-Parking, showcasing the configurations for four EV demand scenarios (50, 100, 150, 200 EVs). \newline
\textit{Note}: CS = Charging Station; CPs = Charging Ports; C = C-Parking; J = J-Parking.}
\label{tab:Jpaking_res}
\scriptsize
\begin{tabular}{lcccc}
\toprule
\textbf{} & \textbf{50 EVs} & \textbf{100 EVs} & \textbf{150 EVs} & \textbf{200 EVs} \\
\midrule
\textcolor{red}{\textbf{CS Configuration}} \\
No. CPs 11kW (C)  & 32  & 50  & 20  & 20  \\
No. CPs 30kW (C)  & 3   & 3   & 10  & 10  \\
No. CPs 11kW (J)  & 15  & 15  & 15  & 15  \\
No. CPs 30kW (J)  & 2   & 8   & 8   & 8   \\
No. Solar Panels  & 604 & 900 & 543 & 600 \\
\midrule
\textbf{Evaluation Metrics} \\
Satisfaction Score  & 1.00 & 0.97 & 0.93 & 0.95 \\
Self-Sufficiency    & 0.75 & 0.61 & 0.56 & 0.57 \\
Normalized Payback  & 0.86 & 0.97 & 0.90 & 0.93 \\
Objective Function  & 2.46 & 2.43 & 2.28 & 2.34 \\
\textcolor{red}{\textbf{No. Cases Tested}} & 52 & 54 & 51 & 53 \\
\bottomrule
\end{tabular}
\end{table}

\section{Discussion}
\label{sec5}
This study presents a Digital Twin framework that integrates dynamic visualization, simulation, and optimization to enable data-driven decision-making for EV charging infrastructure. Implemented on a university campus in Hanoi, the framework facilitates scenario-based evaluations of policies, infrastructure configurations, and renewable energy integration. To further clarify the scope of this framework, it is essential to distinguish between operational and planning-focused digital twins. While operational digital twins prioritize continuous physical-to-virtual synchronization for live system control, our proposed framework is a planning-focused digital twin designed for interactive scenario evaluation and strategic foresight. Its primary value lies in its role as a “virtual laboratory”, enabling planners to test complex policy-infrastructure configurations and explore long-term optimization strategies without the constraints of live data streams or immediate operational requirements. While developed at a localized scale, the model’s modular agent-based design and spatial flexibility allow seamless scaling to larger urban contexts such as innovation districts or emerging urban areas, where the expansion of EV infrastructure presents significant planning challenges. This provides a transferable template for integrating networked multi-station EV charging network across city zones without fundamental changes to the agent logic.

The interactive dashboard supports decision-making by offering dynamic visualizations, performance tracking, and scenario simulations. Our digital twin enables seasonal analysis, revealing a 21\% decrease in solar efficiency from October to March and minimal wind contribution ($\leq5$\% of demand), despite a 28\% investment increase. While hybrid systems enhance grid performance, our findings indicate that in low-wind regions like Hanoi, peak wind generation often occurs outside charging hours (7:00 PM–6:00 AM), limiting effectiveness. Optimizing BESS capacity and demand-response strategies becomes critical for intermittency management. Future research could expand on this by exploring dynamic pricing scenarios, such as peak-hour pricing, where optimizing BESS operations becomes crucial for enhancing the system's ability to integrate solar energy effectively.

The proposed Digital Twin framework enables stakeholders to simulate, evaluate, and optimize EV charging configurations before deployment, ensuring efficient, sustainable networks. Our simulation shows that notification systems improve satisfaction by 10\% during high demand, while gasoline bans and idle fees increase turnover. However, policy interventions alone are insufficient for long-term planning, highlighting the need for expanded capacity. The embedded optimization framework minimizes computational overhead, ensuring near-optimal configurations for fast (30kW) and standard (11kW) chargers integrated with solar energy. Although agent-based modeling and optimization algorithms are well-established methods, this research advances the state of practice by embedding metaheuristic optimization directly into a dynamic simulation platform. This integration eliminates the traditional separation between simulation and post-simulation optimization phases, resulting in a unified framework with an interactive dashboard that is capable of evaluating infrastructure performance, user satisfaction, and energy outcomes in a single workflow. This methodological contribution enhances dynamic adaptability, minimizes computational effort by over 80\%, and provides a unified framework for exploring operational policies and infrastructure strategies simultaneously.

The model has certain limitations, particularly regarding dataset assumptions and the representativeness of the initial local survey. However, the modular design of the agent-based framework allows for flexible integration of stochastic and site-specific behavioral data without altering the model architecture. To strengthen empirical realism, a follow-up one-month field observation and survey (October–November 2025) was conducted to validate behavioral parameters and key assumptions. The results confirmed consistent weekday mobility and charging patterns for parameter setup, with observed increases in both the number of charging stations and EVs already covered within the simulated scenario range, thereby strengthening the robustness of the initial behavioral setup. These updated datasets, included in the Appendix in GitHub repository, demonstrate the framework’s adaptability to evolving user behavior and local EV adoption trends. 

Focusing on a single site also constrains analysis of inter-station interactions across broader networks. However, its modular architecture and GIS-based environment layer allow direct substitution with larger spatial and infrastructure datasets, and the embedded optimization routines and agent logic can be applied at each node without core logic changes, requiring only tuning of probabilistic parameters. In future work for a city-scale multi-node extended project, hierarchical or multi-level decomposition could be adopted, where each charging site or urban sub-zone functions as an independent sub-model exchanging aggregated information with a higher-level coordination layer. This structure supports parallel computation, enabling efficient distribution of agent processes across multiple cores or clusters (as demonstrated in frameworks such as Repast HPC, Flame GPU, etc.), and thereby facilitating city-wide scalability while preserving behavioral realism. Extending the framework to district or city levels requires three practical additions: (1) inter-node interaction modules to simulate vehicle rerouting, multi-purpose trip chains, and station-choice behavior across zones, thereby capturing demand spillover and queue migration dynamics more realistically; (2) coupling with distribution-grid constraint models or powerflow solvers to capture transformer loading, feeder limits, and voltage constraints; (3) richer stochastic demand synthesis and behavioral heterogeneity; and (4) computational strategies (hierarchical optimization, surrogate models, or parallel execution) to manage increased complexity. We have published modular code and implementation guidance to facilitate adaptation of the single-station model to other spatial context deployments. In conclusion, our proposed framework aims to enable the digital twin to support robust, data-driven planning of scalable, resilient EV charging networks.

\section*{Acknowledgement}
This work was supported by the Center for Environmental Intelligence (CEI) at VinUniversity under the research project “Digital Twin Platform to Empower Communities towards an Eco-friendly and Healthy Future” (Project Code: VUNI.CEI.FS\_0001).

\bibliography{bibliography}

\end{document}